\documentclass[conference]{IEEEtran}
\IEEEoverridecommandlockouts
\usepackage{cite}
\usepackage{amsmath,amssymb,amsfonts}
\usepackage{algorithmic}
\usepackage{hyperref}
\usepackage{graphicx}
\usepackage{textcomp}
\usepackage{xcolor}
\usepackage{tabularx}
\usepackage{balance}
\usepackage{caption,subcaption}
\def\BibTeX{{\rm B\kern-.05em{\sc i\kern-.025em b}\kern-.08em  T\kern-.1667em\lower.7ex\hbox{E}\kern-.125emX}}
\begin{document}

\title{
DQC$^2$O: Distributed Quantum Computing for Collaborative Optimization in Future Networks
}
\author{
    \IEEEauthorblockN{Napat Ngoenriang, Minrui Xu,  Jiawen Kang, Dusit Niyato, Han Yu, and Xuemin (Sherman) Shen
    }

\thanks{N.~Ngoenriang is with the School of Information Science and Technology, Vidyasirimedhi Institute of Science and Technology, Thailand (e-mail: \href{mailto:naphat.n\_ s17@vistec.ac.th}{naphat.n\_s17@vistec.ac.th}.}
\thanks{M.~Xu,~H.~Yu~and~D.~Niyato are with the School of Computer Science and Engineering, Nanyang Technological University, Singapore (e-mail: \href{mailto:minrui001@e.ntu.edu.sg}{minrui001@e.ntu.edu.sg}; \href{mailto:han.yu@ntu.edu.sg}{han.yu@ntu.edu.sg}); \href{mailto:dniyato@ntu.edu.sg}{dniyato@ntu.edu.sg}).}
\thanks{J.~Kang  is with the School of Automation, Guangdong University of
Technology, China (e-mail: \href{mailto:kavinkang@gdut.edu.cn}{e-mail: kavinkang@gdut.edu.cn}.}
\thanks{X.~(Sherman)~Shen is with the Department of Electrical and Computer Engineering, University of Waterloo, Waterloo, ON, Canada, N2L 3G1 (e-mail: \href{mailto:sshen@uwaterloo.ca}{sshen@uwaterloo.ca}).}
}
\maketitle
\begin{abstract}

With the advantages of high-speed parallel processing, quantum computers can efficiently solve large-scale complex optimization problems in future networks. However, due to the uncertain qubit fidelity and quantum channel noise, distributed quantum computing which relies on quantum networks connected through entanglement faces a lot of challenges for exchanging information across quantum computers. In this paper, we propose an adaptive distributed quantum computing approach to manage quantum computers and quantum channels for solving optimization tasks in future networks. Firstly, we describe the fundamentals of quantum computing and its distributed concept in quantum networks. Secondly, to address the uncertainty of future demands of collaborative optimization tasks and instability over quantum networks, we propose a quantum resource allocation scheme based on stochastic programming for minimizing quantum resource consumption. Finally, based on the proposed approach, we discuss the potential applications for collaborative optimization in future networks, such as smart grid management, IoT cooperation, and UAV trajectory planning. Promising research directions that can lead to the design and implementation of future distributed quantum computing frameworks are also highlighted.
\end{abstract}


\section{Introduction}
In the quantum era, the advancement of quantum computing and communication has attracted significant interest from researchers, government organizations and the industry~\cite{9684555}. 
Quantum computers provide effective solutions to complex optimization problems in a resource efficient manner, a feat not possible for classical computers to achieve. The recent adoption of quantum computing has further boosted technology innovations such as artificial intelligence (AI), intelligent traffic monitoring, weather forecasting, improved battery chemistry, and life-saving pharmaceuticals~\cite{9684555}.
The proliferation of quantum computing has also evolved to accelerate computation for various applications and services in future networks, which must be designed efficiently using limited resources to perform a wide range of heterogeneous tasks~\cite{bhatia2020quantum}.

The principles of quantum mechanics are used to enable quantum bits (i.e., qubits) in quantum computers which are superior to classical computers based on binary bits.
For example, Shor's \cite{yimsiriwattana2004distributed} algorithm and  Grover's~\cite{qiu2022distributed} algorithm, two well-known quantum algorithms, were developed to efficiently solve factorization and search for unstructured data, respectively, These tasks are highly challenging for classical computers. 
However, scaling up quantum computers is a key challenge in deploying quantum computers in practice. To date, only a small number of qubits can be implemented in a single quantum computer. Moreover, quantum tasks usually require the use of multiple qubits in more complex applications. Due to the instability of qubits and the amount of information required, it becomes more difficult to manage and control information in quantum computers with a small number of qubits.
Thus, distributed quantum computing has been proposed in an attempt to alleviate this problem~\cite{wehner2018quantum}.

The concept of distributed quantum computing, which refers to multiple interconnected quantum computers, was introduced to accelerate and perform collaboratively quantum computations through quantum networks. 
However, due to the principles of quantum mechanics, qubits cannot be duplicated or cloned. Distributed quantum computing requires quantum teleportation, in which qubits are teleported between two quantum computers. 
In this way, a large, complex computational task can be accomplished jointly by multiple quantum computers. 
It has been shown that the most commonly used quantum algorithms benefit from their distributed counterparts.
For example, distributed Grover's algorithm~\cite{qiu2022distributed} incurs significantly shorter query time than Grover's algorithm, and distributed Shor's algorithm \cite{yimsiriwattana2004distributed} is less complex than Shor's algorithm. The distributed versions of both algorithms increase their viability of solving complex problems in practice.

\begin{figure*}[t]
    \centering \includegraphics[width=1\linewidth]{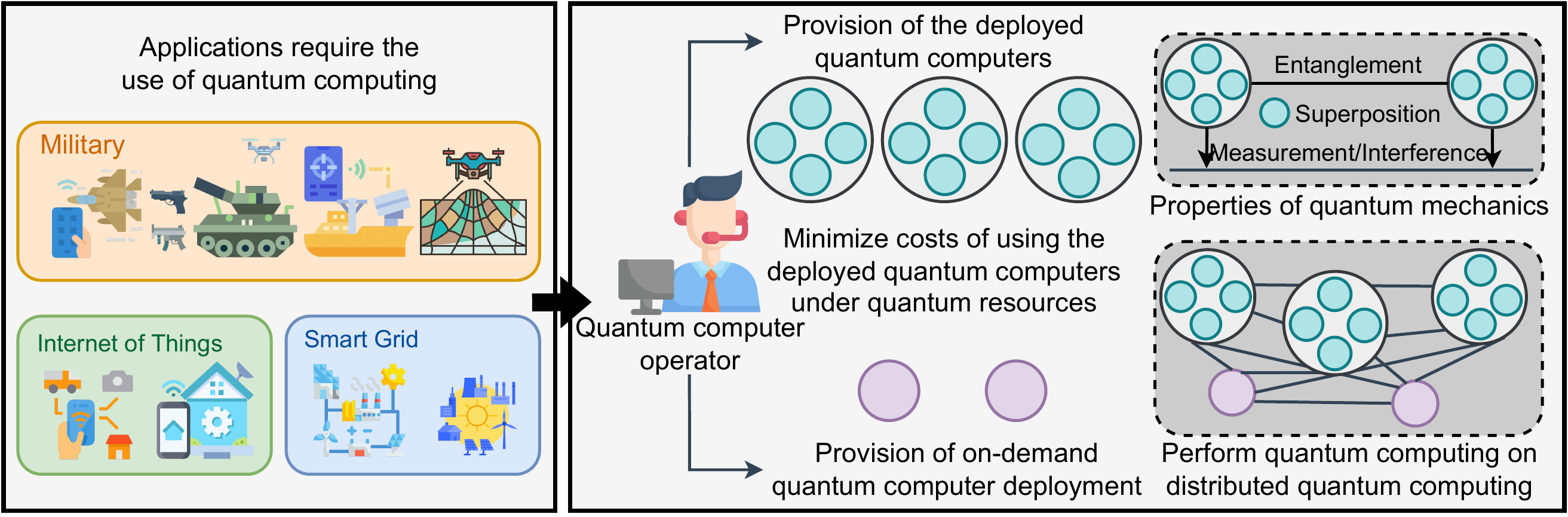}
    \caption{The adaptive resource allocation approach with two-stage stochastic programming for distributed quantum computing.}
    \label{fig:timeline}
\end{figure*}

Although distributed quantum computing has many advantages over a single quantum computer and classical computers, designing efficient large-scale distributed quantum computing still faces many challenges. 
Firstly, the effectiveness of using quantum computers is determined by the demand of the applications, like military communication. This is unknown at the time of deployment, making planning difficult. 
Secondly, the availability and computing power of quantum computers, which may vary over time, also affects whether quantum tasks can be fully computed.
Thirdly, distributed quantum computing may suffer from fidelity degradation. This is unavoidable at the moment, and reduces the efficiency of quantum teleportation in quantum networks.
Thus, deploying quantum resources in distributed quantum computing in its current form may result in highly inefficient utilization of quantum computing resources due to the inherent uncertainty in real-world circumstances. 

To address the above challenges, in this paper, we propose an adaptive resource allocation approach towards efficient and scalable distributed collaborative quantum computing. It consists of deterministic and stochastic programming models for quantum resource allocation with uncertainty in collaborative settings to help quantum computer operators minimize total deployment costs.
It jointly considers the uncertainty of future quantum computing demands, the computing power of quantum computers, and the fidelity in distributed quantum computing in order to optimally deploy and utilize quantum computers.
We conduct extensive experiments to reveal the importance of the optimal deployment of quantum computers in distributed quantum computing. In comparison to other resource allocation models, the proposed approach can reach the lowest total deployment cost. 
Finally, we highlight opportunities and challenges in distributed quantum computing for various military applications in future networks.

\begin{center}
\begin{table*}[t]
\small
\begin{tabularx}{\textwidth}{|c|X|X|X|X|}
\hline
\textbf{Ref.}&\textbf{Scenarios}&\textbf{Problems}& \textbf{Performance Metrics}& \textbf{Quantum Algorithms}\\
\hline
\cite{eskandarpour2020quantum} & Quantum-enhanced smart grid & Paradigm and elements of quantum computing in power systems & Grid efficiency, security, and stability & Quantum Fourier transform, quantum search, and quantum neural networks \\\hline
\cite{bhatia2020quantum} & Realizing IoT environments & IoT-sensor space problem & Data accuracy and data temporary efficiency & Quantum-based algorithm and quantum Fourier transform\\\hline
\cite{li2021intelligent}& UAV-mounted wireless networks & Trajectory and uplink transmission rate optimization & Convergence performance and learning quality of trajectories & Quantum-based action selection strategy \\\hline
\cite{mlquantum}& Quantum machine learning & Improvement computation of machine learning algorithms & Computation complexity & Quantum mechanics, Shor's, and Grover's algorithm\\\hline
\cite{Lee2022router}& Entanglement flow in quantum networks & Maximization of fidelity over long-distance links & Entanglement rate and router infidelity & Quantum router architecture design\\\hline
\cite{Tiersch2015adap}& Adaptive measurements in quantum & Estimation state of qubits & Success probability of estimation & Grover's algorithm\\\hline
\cite{cicconetti2022resource} & Resource allocation in quantum networks & Quantum network flow optimization & The number of served apps, the Bell pair capacity, and Jain's fairness & (Weighted round robin algorithm for resource allocation)\\
\hline
\end{tabularx}
\hfill{}
\caption{Summary of representative related works demonstrated through scenarios, problems, performance metrics, and quantum algorithms}
\label{tb:summary}
\end{table*}
\end{center}

\section{Fundamentals of Quantum Computing}

\subsection{Quantum Computing}
Three properties of quantum mechanics  define quantum computing: superposition, interference, and entanglement, as shown in Fig. \ref{fig:timeline}. 

\subsubsection{Superposition} 
In classical computers, the binary bits 0 and 1 are used to encode information for computations. A superposition in quantum computing allows the encoding of qubits in a combination of two classical binary states. 
A well-known example of quantum mechanics is the tossing of a coin. When a coin is tossed in the air, the exact outcome is unknown, and its probability values reflect qubits.
Each qubit can be expressed as a linear combination of binary states, where the coefficients correspond to the probabilities of the qubit amplitudes.
Due to the superposition, $n$ qubits can store $2^n$ possible outcomes and have the same chance of being measured for each. Therefore, quantum computers can store and manipulate more information than classical computers, providing far more diverse possibilities and opportunities.

\subsubsection{Interference} 
Qubits must be subjected to some kind of measurement in order to represent and store their values and results. If one intervenes in the process, it is possible to measure and see the results of the paths. When the process is interrupted, the states of the qubits collapse to classical bits, and the computation results appear. For example, the result of a coin toss is known definitively (e.g., heads or tails) when the coin reaches the bottom.

\subsubsection{Entanglement} 
Two qubits can be entangled with each other as an entanglement pair~\cite{wehner2018quantum}, which means that when one qubit is measured, the other qubit can also be known because of their entangled nature. In addition, a pair of entanglement qubits is entangled maximally also known as a Bell state when the results of measuring one of them will certainly affect the outcome of measuring the other one later.
The fidelity of entanglement pairs is the metric of attenuation for the entangled qubits between two remote quantum computers.
The fidelity scale runs from 0 to 1, where 1 indicates the best performance that the entanglement can achieve.

The main analogies between classical computing and the technology used to realize a true quantum computer are the following. In classical computing, a circuit is a computational model that enables the processing of input values through gates and operations. Similarly, the \textit{quantum circuit model} proceeds with implementations on qubits and involves an ordered sequence of \textit{quantum gates} that permit logical interaction between qubits. In particular, the measurement of qubits must occur near the end of the quantum circuit. A \textit{quantum processor} is a small quantum computer that can execute quantum gates on a small number of qubits and allows the entanglement of qubits inside.

\subsection{Distributed Quantum Computing}
The concept of distributed quantum computing relies on the following principles:

\subsubsection{Quantum Networks through Entanglement} 
    To enable distributed quantum computing, a quantum network must be developed to connect quantum computers.
    However, due to the non-cloning theorem in quantum mechanics, qubits cannot be duplicated or cloned across quantum computers. 
    Thus, two quantum computers must exchange or transfer qubits through the concept of quantum teleportation.
    In quantum teleportation, a pair of qubits from the source quantum computer to the destination quantum computer is transferred for local operations and measurements before decoherence, the loss of the shared entangled qubits.
    However, errors and failures incurred in quantum teleportation preserve the entangled qubits of quantum computers.
    To guarantee the Bell state of the shared pair of qubits, an entanglement distillation procedure is developed to iteratively increase fidelity or the probability that the qubit will be transferred without changing its state. However, quantum teleportation requires a lot of iterations, time, and resources to achieve the Bell state of a pair of qubits.
    Additionally, the qubits of the shared Bell pairs are called entangled-qubits (or eqbits), which are stored in quantum memories.
    
\subsubsection{Distributed Quantum Circuits} 
    Required quantum circuits can be split and distributed among multiple quantum processors of quantum computers, with each quantum computer executing a fragment of quantum circuits. The architecture of the quantum circuits executes non-local quantum gates for the shared entangled qubits. A logically identical set of instructions is used to coordinate the additional operations required for the non-local operation to replace the multiple qubit operations. In particular, a partitioning of quantum circuits to achieve the least number of qubits among quantum processors must be determined remotely by exchanging non-local operations with other quantum gates.
    
\subsubsection{Distributed Quantum Architectures} 
    Multiple quantum processing units (QPUs) can be executed for computations in parallel with each other universally to achieve distributed quantum computing. Qubits are prepared as input data by the QPUs, which are subsequently utilized to read output after a measurement. 
    The QPUs should have specialized space to store and operate matter qubits that improve the ability of distributed quantum computing to work quickly and consistently. In addition, network communication is also needed for QPUs to send signals across the network while measuring and reading qubits.
    The effort required to input data and the quality of matter qubits is a trade-off. 
    Therefore, the quantum architectures, including quantum algorithms, need to be well-designed and well-resolved to achieve the communication and local operations for generating matter qubits.

\subsection{Quantum Algorithms and Distributed Quantum Algorithms} 
The quantum algorithms to be executed in quantum computers can be of several types depending on the computational paradigm. The key issues on quantum algorithms and distributed quantum algorithms are as follows.

\subsubsection{Quantum Algorithm}
Numerous studies have focused on quantum algorithms operating at exponentially faster speeds on quantum computers implemented on quantum circuits~\cite{quannetbook}. We discuss the most prevalent quantum algorithms as follows:
\begin{itemize}
    \item Shor's algorithm was devised to factor prime numbers in polynomial time, which cannot be done polynomially by classical algorithms. The factoring problem is suggested to be reformulated as the finding-period problem and solved through quantum phase estimation, which estimates phases or eigenvalues of eigenvectors of unitary operators.
    \item Grover's algorithm solves a searching problem for an unstructured database, which provides quadratically speed-up computation. An equal superposition of all possible solutions is employed as inputs that result in the same amplitudes. The inputs are passed and processed through the oracle diffuser to boost the amplitudes and then reflect the solution in terms of the greatest amount of amplitudes.
\end{itemize}
The Grover's algorithm can be used as a generic algorithm to address a variety of problems due to the independence of the algorithm and the internal structure of lists. As a result, executing Grover's algorithm gives many classical problems a quadratic speed-up computation.


\subsubsection{Distributed Quantum Algorithm}
\begin{figure}
  \centering
  \begin{subfigure}{\linewidth}
  \centering \includegraphics[width=0.8\linewidth,height=0.45\linewidth]{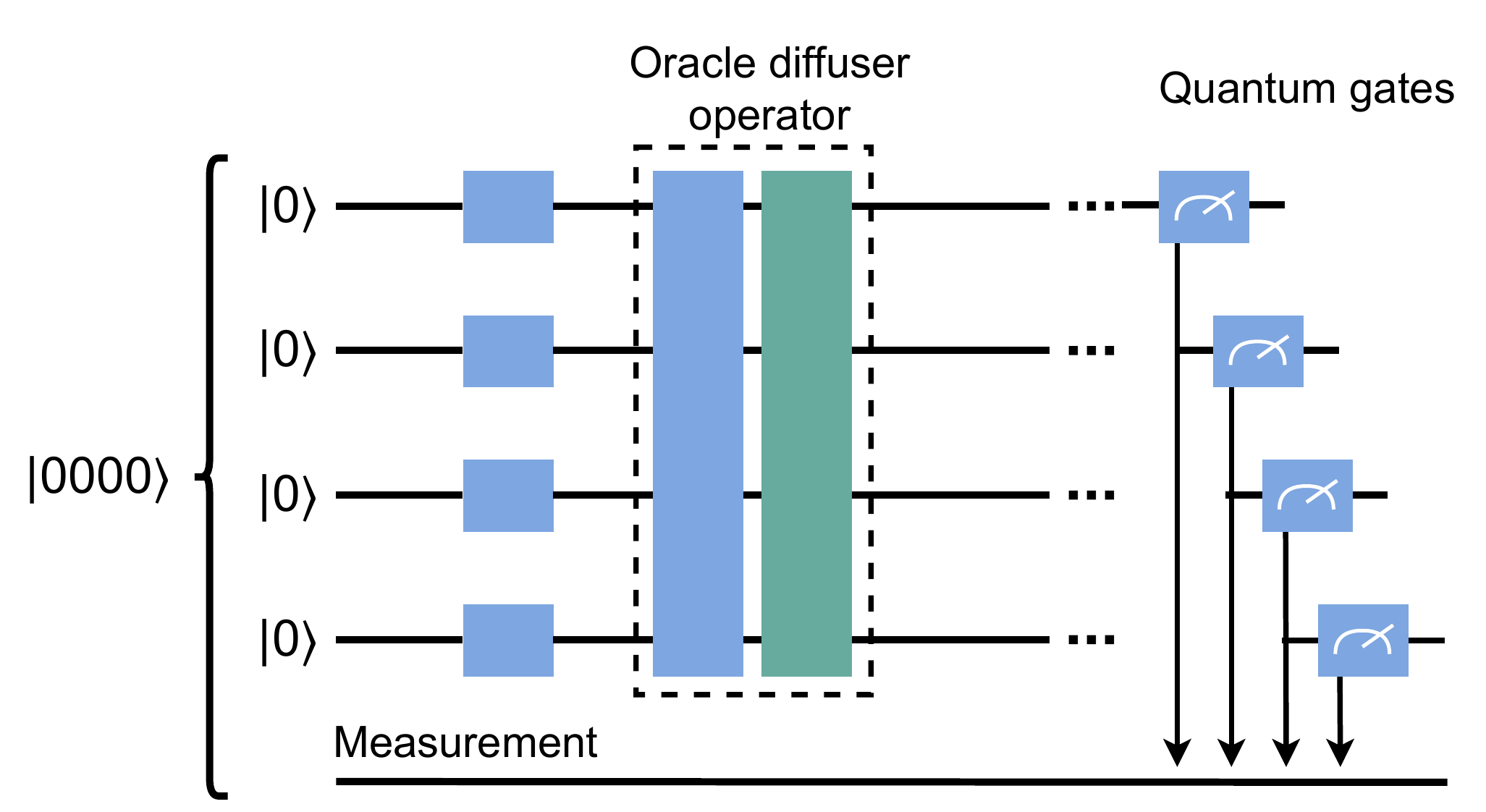}
    \caption{\small}
  \end{subfigure}\\
  \begin{subfigure}{\linewidth}\centering\includegraphics[width=0.8\linewidth,height=0.45\linewidth]{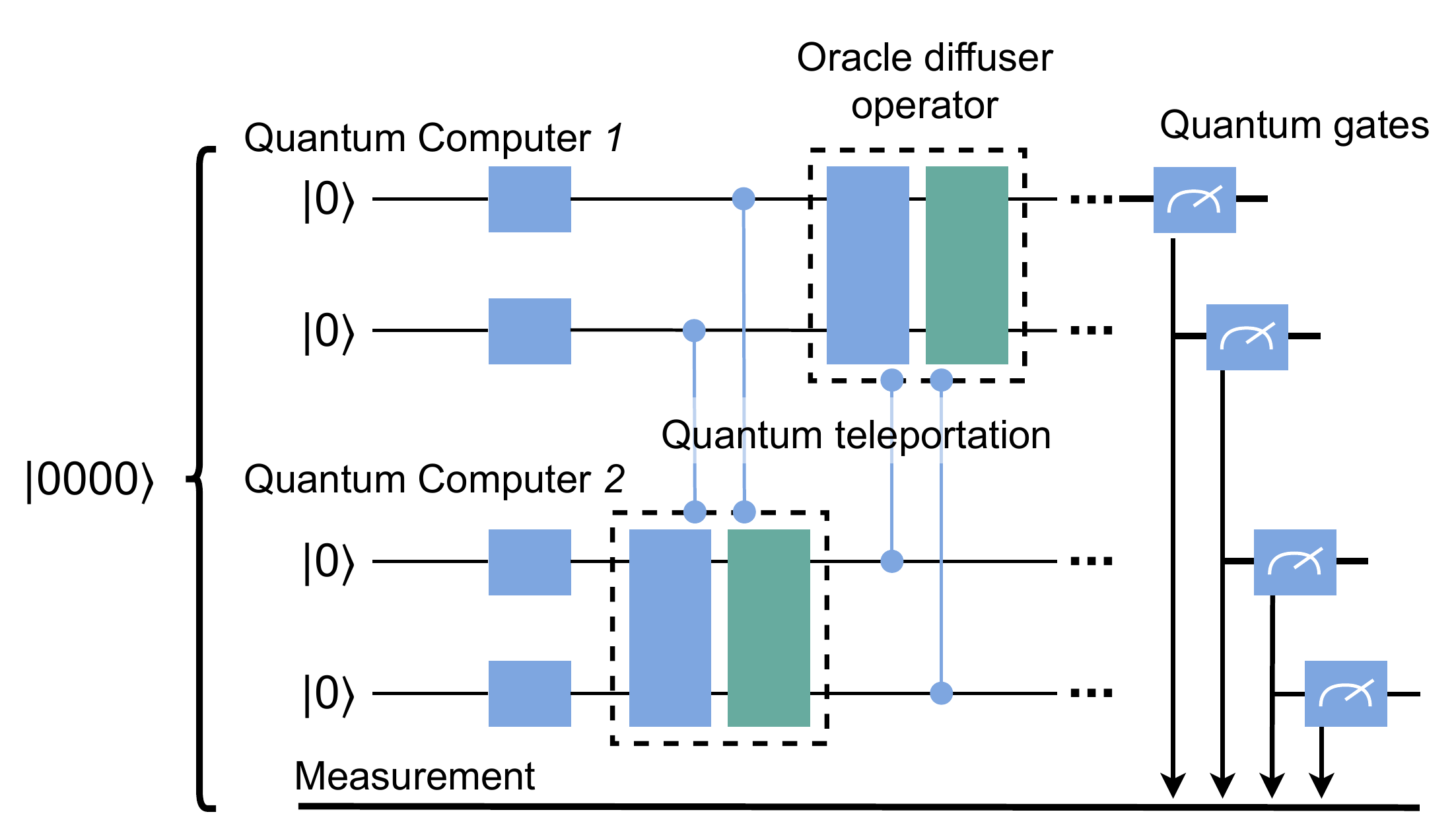}
    \caption{\small}
  \end{subfigure}
  \caption{A procedure of computing the required four-qubits quantum task; (a) one quantum computer with four qubits and (b) two quantum computers with two qubits per each.} \label{fig:dqc}
\end{figure}

Despite the fact that quantum computing is significantly more advanced than classical computing, only small quantum computers (with a limited number of qubits) have so far been constructed due to the noise and depth of quantum circuits. As an alternative to large-scale monolithic designs, a distributed grid of small quantum computers has been proposed to help advance quantum computing.
In particular, as distributed quantum computing necessitates sharing the superposition state among quantum computers, quantum networks through entanglement need to be considered in the implementation. In addition, swap gates are implemented in order to teleport qubits between a pair of two quantum computers. We discuss the most prevalent distributed quantum algorithms as follows:
\begin{itemize}
    \item Distributed Shor's algorithm was implemented to solve factorization problems by small-capacity quantum computers. The architecture of quantum circuits to perform collaboratively is designed to handle quantum teleportation and simulate a large capacity quantum computer \cite{yimsiriwattana2004distributed}. The complexity of distributed Shor’s algorithm is $O\!((\log N)^2)$, while Shor's algorithm requires $\displaystyle O\!\left((\log N)^{2}(\log \log N)(\log \log \log N)\right)$, where $N$ denotes the integer to be factored.
    \item Distributed Grover's algorithm was also developed to address the unstructured search problem. According to the expensive query time in Grover's algorithm, the distributed Grover's algorithm can reduce query times \cite{qiu2022distributed}. Functions that needs to be computed can be divided into $2^k$ subfunctions by the distributed Grover's algorithm, which then computes one of the usable subfunctions to find the solution to the original function. In comparison to the original Grover's algorithm, distributed Grover's algorithm can significant speed up queries.
A procedure of Grover's algorithm solving a four-qubit quantum task for quantum computing and distributed quantum computing using two quantum computers is shown in Fig.~\ref{fig:dqc}.
\end{itemize}

Among distributed quantum algorithms, the grid of distributed quantum circuits has to be well-designed to make the best utilization of the constrained resources in quantum computers. 
Therefore, it becomes increasingly important to address resource allocation in distributed quantum computing to solve resource constraints in quantum computing environments.

\section{A Case Study: Resource Allocation in Distributed Quantum Computing}


\subsection{Allocating Resources in Distributed Quantum Computing}
In the proposed distributed quantum computing framework, a quantum computer operator serves quantum computation tasks by distributing each of them to one or multiple quantum computers that can work collaboratively to accomplish the quantum computation. Quantum tasks may require the number of qubits, which each quantum computer provides. The links between two interconnected quantum computers enabling quantum teleportation are oriented and have a static capacity. However, the number of quantum computers, the number of qubits of each quantum computer, and the capacity of quantum networks are limited and static. Therefore, the allocation of these resources needs to provide sufficient quantum deployments to support quantum tasks in the distributed quantum computing framework.

\subsection{System Model, Decisions, and Costs}
As shown in Fig. \ref{fig:timeline}, the quantum computer operator chooses between two alternative stages for allocating resources for distributed quantum computing, i.e., utilizing the reserved quantum computers and deploying on-demand quantum computers. To use the reserved quantum computer, the quantum computer operator follows instructions of the distributed quantum computers to complete the quantum task. 
However, the reserved quantum computers may not be sufficient to accomplish a large-scale quantum computational task. Therefore, the quantum computer operator can request the deployment of new quantum computers, or the utilization of on-demand quantum computing may be required. Shortly, quantum computing from multiple organizations can be shared in which one organization can borrow or buy quantum computing time from another organization temporarily. Such a scenario is inspired by the present conventional cloud computing paradigm. For example, Amazon Braket offers on-demand quantum cloud computing services~\cite{qcloud}. For this option, the quantum computer operator installs and configures the on-demand quantum computers to works collaboratively with the deployed quantum computers.
In addition, the cost of using the deployed quantum computers is less than installing the new quantum computer. Utilizing the deployed quantum computer will cost the computing power of the quantum computer and the Bell pair of the two connected quantum computers. However, using an on-demand quantum computer is more expensive. The objective of the quantum computer operator is to satisfy the quantum computing tasks while minimizing the total deployment cost.

\subsection{Uncertainty of Distributed Quantum Computing}
While determining the best resource allocation in distributed quantum computing, uncertainties occur, which can be classified into three types as follows:

\begin{itemize}
    \item First, the actual demands are unknown when the option of deploying the quantum computer is made. Different applications such as minimization, dimensional reduction, and machine learning problems may request various qubits~\cite{machqubits}. For example, up to 30 qubits are used in machine learning problems \cite{machqubits}.
    \item Second, the precise availability of the quantum computer and its qubits is uncertain since they might be reserved for other purposes or because the quantum computer’s backend may not support all of them~\cite{qibm}. For example, only 5 of 10 quantum computers are available to compute quantum tasks.
    \item Third, the fidelity of the entangled qubits is also not known exactly due to the degradation of the entangled qubits in distributed quantum computing~\cite{quannetbook}. For example, if the fidelity is 0.5, it means that 50\% efficiency of the entangled qubits can be achieved.
\end{itemize}
Therefore, an adaptive resource allocation approach is required to efficiently provision quantum computers to tackle quantum computational tasks with dynamic sizes under uncertainty of the computing power of quantum computers and fidelity fidelity of the entangled qubits.

\subsection{The Proposed Approach}
In the deterministic resource allocation for computing quantum tasks in distributed quantum computing, the required demand of quantum tasks, the computational power of quantum computers, and the fidelity of the entangled qubits are exactly known by the quantum computer operator. Therefore, quantum computers can be certainly deployed and the on-demand quantum computer deployment is not necessary.
However, due to the aforementioned challenges and the uncertain environments in distributed quantum computing, the deterministic resource allocation approach is not applicable and the on-demand deployment will be the solution. Therefore, we propose the adaptive distributed quantum computing approach based on the two-stage stochastic programming model. The first stage defines the number of deploying the reserved quantum computers, while the second stage defines the number of installing the on-demand deployment of quantum computers.
The stochastic programming model can be formulated as the minimization of the total cost including two parts, where the first part is the first-stage cost of using the reserved quantum computers and the second part is the expected second-stage cost that consists of the computing power unit and Bell pair costs of the deployed quantum computers including the on-demand quantum computer deployment under the set of the possible realized demand of quantum tasks, computing power of quantum computers, and fidelity of the entangled qubits. The constraints of the stochastic programming model are the selection of the utilized quantum computers, both the deployed and the on-demand quantum computers being able to complete the computational tasks, and the utilization of the computing power under the link capacity, with considering the uncertainty.

\subsection{Experimental Results}
\subsubsection{Parameter Setting}
We consider the system model of quantum computing where the quantum computer operator consists of 10 deployed quantum computers. 
We set the cost values, which is measured in normalized monetary, for using the deployed quantum computer, qubits, and Bell pairs to be 5000, 1000, and 450, respectively.
All quantum computers have 257 qubit capacities and identical costs. The cost and computing power of new quantum computer deployment are $25000$ and $127$, respectively. Costs for both on-demand and quantum computers are determined based on~\cite{9684555},~\cite{qcloud}.
To solve the proposed stochastic model, we consider two scenarios. The first scenario is that the demand of the quantum task is 10, the available computing power of the quantum computers is 127 qubits, and the fidelity of the entangled qubits in quantum networks is 1 (i.e., the best performance). The second scenario is that there is no demand, no availability of qubits, and zero fidelity of the entangled qubits in quantum networks (i.e., the worst performance). We assume the default probability values with 0.8 and 0.2, respectively.

\begin{figure}
\centering
  \begin{subfigure}{0.95\linewidth}
  \centering \includegraphics[width=\linewidth,height=\linewidth,keepaspectratio]{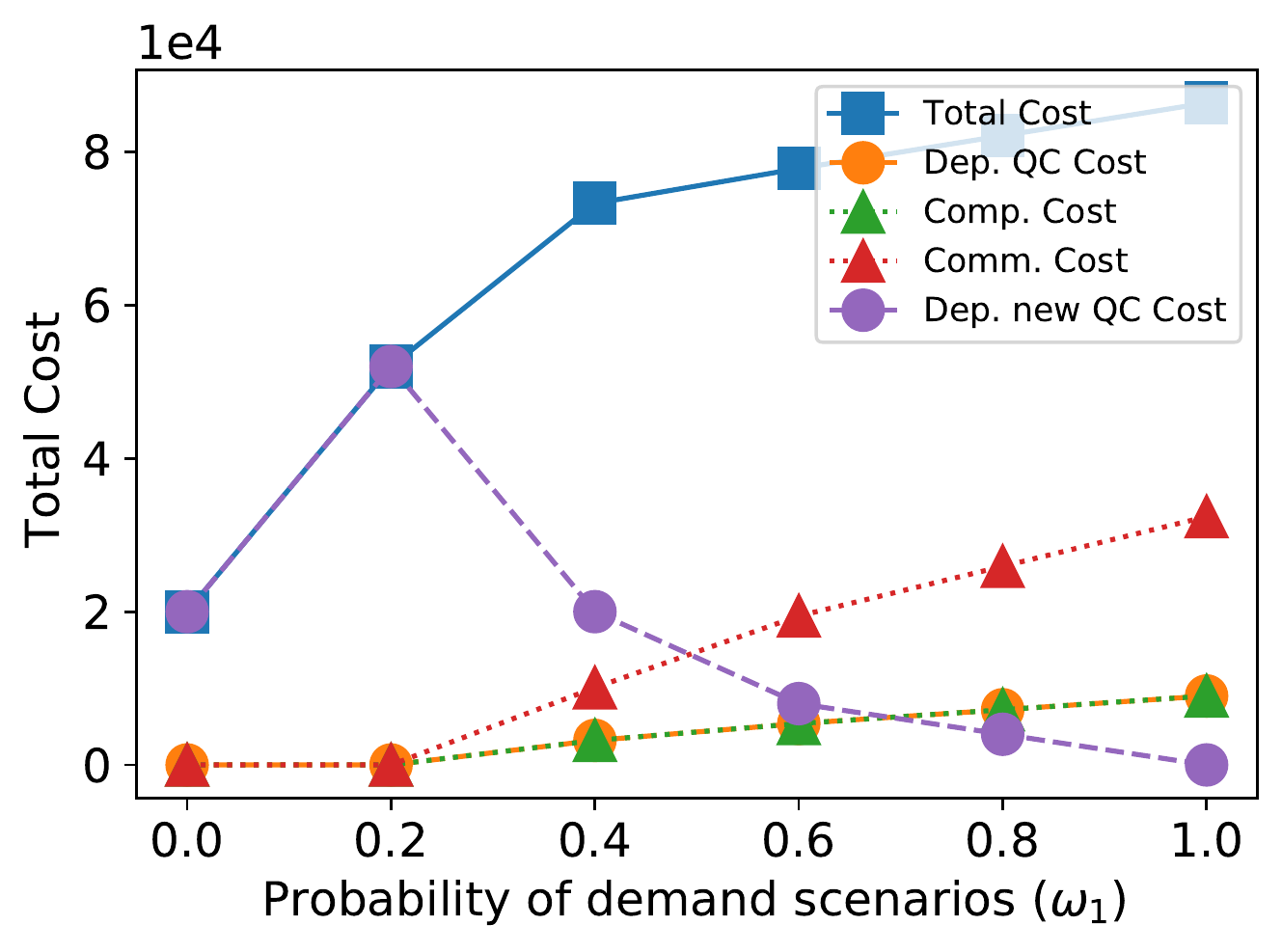}
    \caption{\small Cost breakdown under different probabilities}
    \label{fig:compare}
  \end{subfigure}
  \quad
  \begin{subfigure}{1\linewidth}
    \centering \includegraphics[width=\linewidth,height=\linewidth,keepaspectratio]{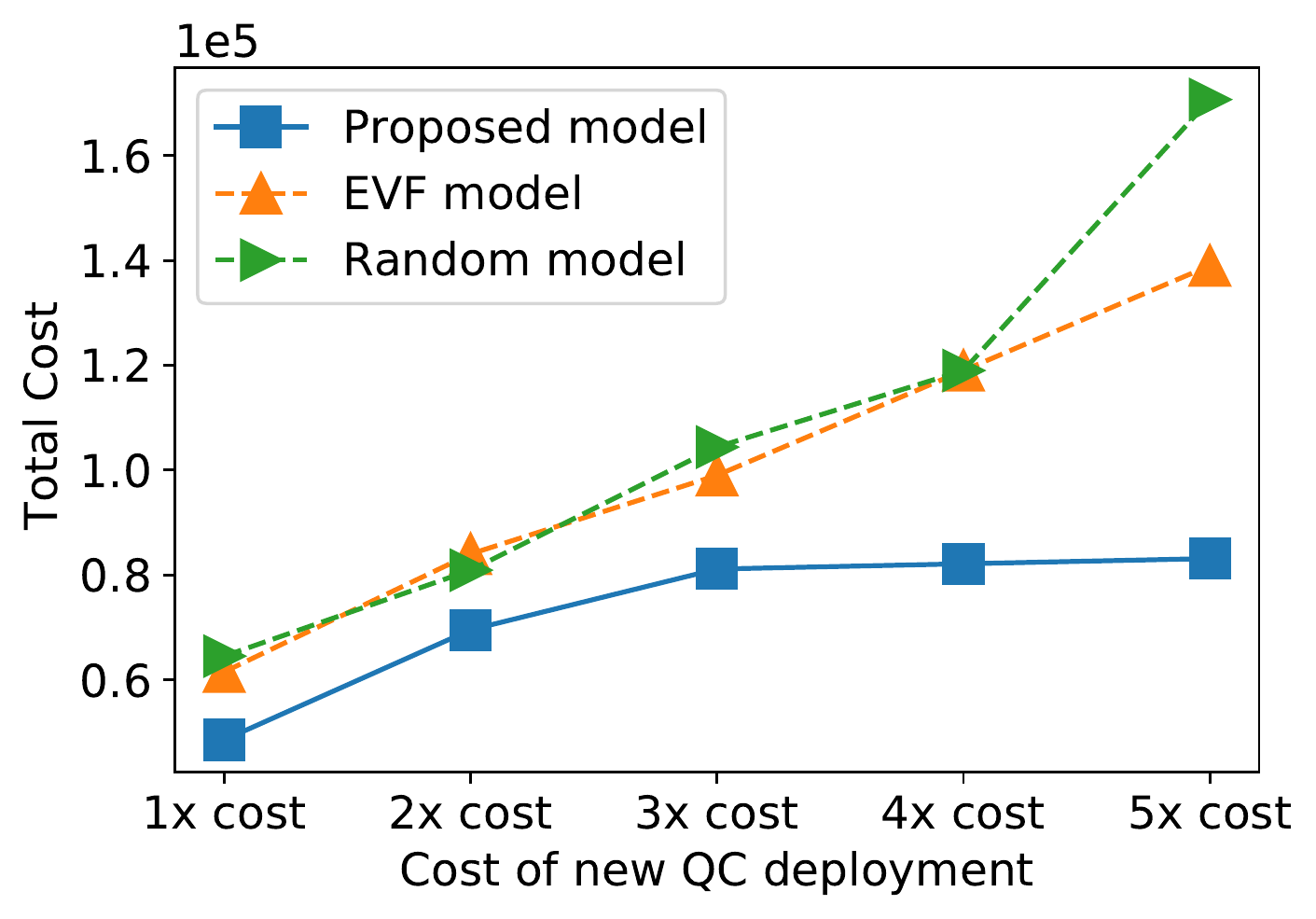}\caption{Comparison among the proposed, EVF, and random models}
    \label{fig:prob}
  \end{subfigure}
  \caption{The cost breakdown and comparison of the stochastic programming model for a resource allocation in distributed quantum computing.}
\end{figure}

\subsubsection{Impact of Probability of Scenarios}
We vary the probability of the first scenario, which corresponds to having the demand, the availability of the quantum task, and the best performance for the entangled qubits. The cost breakdown is shown in Fig. \ref{fig:prob}. We note that when the probability of the scenario is equal to or less than 0.2, the new quantum computer should be deployed. The deployed quantum computer is utilized over the new quantum computer when the probability of the scenario is higher as it has the demand of the quantum task, computing power of quantum computers, and the fidelity.

\subsubsection{Cost Comparison}
We compare the proposed stochastic model with both the Expected Value Formulation (EVF) model and the random model. The EVF model solves the deterministic model using the average values of the uncertain parameters. The cost of the new quantum computer deployment is varied. In the random model, the deployed quantum computer in the first stage is randomly selected. Figure \ref{fig:compare} depicts the comparison of total costs of the three models. We observe that the proposed model achieves the lowest total cost. The minimum total cost cannot be guaranteed by using the average values of the uncertain parameters used in the EVF. In addition, the EVF and random models are unable to account for variations in the deployment costs of the on-demand quantum computers.

\section{Opportunities and Challenges}
\subsection{Applications in Collaborative Optimization}

\subsubsection{Future Smart Grid}

Extending from classical computing, quantum computing can be used to enhance computational approaches that support decision-making in smart grids, as illustrated in~\cite{eskandarpour2020quantum}.
The current smart grid issues also involve power systems that operate on several timescales and dimensions and must be resolved immediately.
To address upcoming smart grid concerns, distributed quantum computing is more feasible and promising than quantum computing.

\subsubsection{Future Internet of Things}

Inspired by the aspects of acquiring Internet of Things (IoT) data and improving  data accuracy analysis, the article~\cite{bhatia2020quantum} presents a novel quantum computing-inspired (IoT-QCiO) optimization to optimize IoT-sensor space by using quantum formalization based on quantum mechanics.
According to the increased amount of IoT-connected devices, the scalability of utilizing one quantum computer needs to be extended to enhance the overall deployment using the properties of distributed quantum computers.


\subsubsection{Future UAV Trajectory Planning}
Optimizing UAVs trajectory is challenging when knowledge of ground users, e.g., locations and channel state information, are unknown.
Quantum-inspired reinforcement learning (QiRL) approach adopts superposition and amplitude amplification in quantum mechanics to select probabilistic action and strategy solved by conventional reinforcement learning on classical computers~\cite{li2021intelligent}. 
The actual quantum computers, including distributed quantum computing, are required to improve convergence speed and learning effectiveness.

These applications may all be categorized as large-scale problems in future networks, which are still challenging because of the numerous computational resources and processes required. A fresh solution to these problems could be further addressed by the distributed quantum computing paradigm. 
Furthermore, classical and quantum computers will still coexist to execute computational tasks.
Hybrid computing, which combines quantum and classical computing, is required to significantly reduce energy consumption and costs.

\subsection{Challenges in Distributed Quantum Computing}

\subsubsection{Efficient Routing for Collaborative Quantum Computers}
With long-distance links in quantum networks, quantum repeaters have been introduced as an intermediate for communicating between quantum computers. 
The quantum router architecture is designed to sustain entanglement over quantum networks consisting of quantum memories coupled through photons \cite{Lee2022router}. 
The quantum router architecture is required to improve entanglement fidelity and deliver effective quantum routing while reducing memory latency across quantum and repeater networks.
Open challenges remain to design an efficient quantum routing to control entanglement fidelity including long-distance links in distributed quantum computing.

\subsubsection{Coexistence of Multiple Distributed Quantum Algorithms}
Since there will be a variety of optimization problems with different structures in future networks, the corresponding quantum algorithms required for them will be different accordingly, from the number of qubits to the type of quantum gates. For a single optimization problem, distributed quantum computing can divide the problem into multiple homogeneous subproblems and deploy them on interconnected quantum computers for cooperative solutions. However, to meet the multi-task and multi-algorithm requirements, distributed quantum computing must be equipped with the ability to handle heterogeneous quantum computing tasks. Therefore, standardized protocols and efficient collaboration will be indispensable in distributed quantum computing in future complex network optimization problems, e.g., critical and secure communication in military environments.

\subsubsection{Adaptive Quantum Resource Allocation}
According to various applications in future networks, such as mission-critical applications in military communications, distributed quantum computing must consider dynamic environments of using quantum resources and channels. 
To measure qubits, there are possibilities on entanglement of qubits that may incur errors. The distributed quantum computing can be designed with collaboratively adaptive measurement directions and deployment across multiple quantum computers.
Further requirements and limitations on quantum settings and resources will be essential for efficient DQC.

\subsubsection{Distributed Quantum Machine Learning}
In contrast to conventional machine learning, quantum machine learning was proposed using quantum algorithms to carry out machine learning more rapidly. Multiple paradigms of the present machine learning algorithms, encoded in a superposition state, are used to improve machine learning solutions~\cite{mlquantum}. For example, quantum principal component analysis (PCA) is exponentially more effective than traditional PCA~\cite{mlquantum}. According to the capability of distributed quantum computing, more exciting issues exist in various machine learning applications, such as quantum federated learning, in which multiple clients (or quantum computers) can work collaboratively and securely with distributed quantum algorithms. Therefore, future networks can benefit from applying quantum machine learning and distributed quantum machine learning to address more complex problems and improve the performance of the classical machine learning.

\section{Conclusions}
In this paper, we present the fundamentals of quantum computing, with special focus on distributed quantum computing. It relies on quantum networks connected through entanglement, and the architecture of distributed quantum circuits to collaboratively perform complex computational tasks.
To efficiently utilize quantum resources at scale and in the face of uncertainties, we have proposed an adaptive resource allocation approach to achieve cost efficiency in distributed quantum computing. The experimental results show that the proposed model can minimize the total cost under uncertain distributed quantum computing environments. Finally, we have discussed the opportunities and challenges of distributed quantum computing in future networks.

\balance
\bibliographystyle{IEEEtran}
\bibliography{ref}
\end{document}